# Ultra-wideband electrostrictive mechanical antenna

**Jianchun Xu[1†], Zhao Li[1,2†], Xuchao Pan[3†], Xi Wen[3], Jinqing Cao[1], Wen Gong[4], Shaolong Yang[1], Ming Lei[1★], Fangzhou Yao[5★] and Ke Bi[1★]**

[1]*State Key Laboratory of Information Photonics and Optical Communications, School of Electronic Engineering, School of Science, Beijing University of Posts and Telecommunications, Beijing 100876, China.*
[2]*State Key Laboratory of New Ceramics and Fine Processing, School of Materials Science and Engineering, Tsinghua University, Beijing 100084, China.*
[3]*School of Mechanical Engineering, Nanjing University of Science and Technology, Nanjing 210094, China.*
[4]*Research Center of Advanced Functional Ceramic Materials, Wuzhen Laboratory, Jiaxing, 314500, China.*
[5]*Center of Advanced Ceramic Materials and Devices, Yangtze Delta Region Institute of Tsinghua University, Zhejiang 314006, China.*

[†]These authors contributed equally to this work.
[★]e-mail: mlei@bupt.edu.cn; yaofangzhou@tsinghua-zj.edu.cn; bike@bupt.edu.cn.

**Conventional mechanical antennas provide a strategy in long-wave communication with a surprisingly compact size below 1/1,000 of the wavelength. However, the narrow bandwidth and weak field intensity seriously hamper its practical applications. Here, we present a mechanical antenna based on the electrostrictive effect of PMN-PT-based relaxor ferroelectric ceramic to improve radiation capacity and achieve ultra-wideband characteristics (10 kHz - 1 MHz, the relative bandwidth is beyond 196%). Determined by the different underlying mechanism, the mechanical antenna based on the electrostrictive effect exhibits excellent communication properties from traditional mechanical antennas. The functions of signal coding, transmitting, receiving, and decoding were experimentally demonstrated. This approach offers a promising way of constructing mechanical antennas for long-wave communication.**

Electromagnetic radiation at very low frequency (VLF, 3 - 30 kHz) exhibits relatively low propagation loss, enabling remote communication[1,2]. Electrically small antennas are of both practical and fundamental importance in portable wireless devices due to the advantage of miniaturization[3]. According to Chu limit, the minimum radiation quality factor of an electrically small antenna is limited by its size[4]. Due to the large reactance, these antennas generally have narrow bandwidth. Various shapes, high-contrast materials, metamaterials, and lumped components have been employed for improving their bandwidth and radiation resistance[5,6]. However, even the electrically small antennas have a large physical size. They also need an additional impedance matching network in the VLF band, which is not beneficial for practical applications in long-wave communication. Recently, a new mechanical antenna was proposed and attracted extensive attentions[7,8].

Mechanical antenna based on mechanical motion is a miniaturization antenna, its physical size can be smaller than 1/1,000 of the operating wavelength. Conventional electrically small antennas rely on oscillating charges to radiate electromagnetic (EM) waves, while mechanical antennas are driven by acoustic self-resonance to generate the mechanical motion of charge or magnetic dipole moment oscillations, thus generating time-varying EM field[9]. Because of its self-resonance at acoustic wavelength, such mechanical antennas can eliminate the need for a bulky, external impedance-matching network. According to the latest reports, piezoelectric materials, such as quartz, single crystal lithium niobate (LN), and lead zirconate titanate (PZT), have been successfully used to generate acoustic resonance for radiating EM waves in VLF band[10]. However, their polarized state is difficult to maintain in the extreme working environment of high temperature and high pressure. In addition, mechanical antennas based on piezoelectric materials usually have a narrow bandwidth due to their high quality factor[11-13]. For example, an LN piezoelectric electric dipole presented for high radiation efficiency can only operate near 35.5 kHz[14]. In many civil fields and military areas, wideband is urgently desired to provide several important advantages including high date rate, resistances to interference, covert transmission, and so on. Such advantages enable a wide range of applications of communications, radar, imaging, and positioning, which obviously cannot be offered by the narrow band. It is therefore important to propose a new radiation mechanism for wideband mechanical antennas.

Here, we present a set of strategy and design concept to expand the bandwidth of mechanical antenna. Based on the electrostrictive effect of dielectric material, ceramic cylindrical resonator is expected to radiate EM waves under a periodic electric field. The detailed discussion about the difference between the electrostrictive effect and the inverse piezoelectric effect in mechanical antennas application can be found in Supplementary Information. The operating mechanism and ultra-band characteristics distinguish our work from the previously reported strategies. In this work, mechanical antennas based on the electrostrictive effect were fabricated using 2.5Sm-PMN-29PT relaxor ferroelectric ceramics. The ceramics cylindrical resonator showed superior property to radiate EM waves under a moderate periodic electric field. The feasibility of this design scheme was verified by demonstration experiments of the basic communication functions, *i.e.*, signal modulation, transmission, and receiving within an ultra-band. It's believed that the characteristics of ultra-band and high radiation efficiency of this mechanical antenna must accelerate its practical process in long-wave communication.

## Results

For a mechanical antenna, the signals are directly fed into the resonator without external matching network (Fig. 1a). According to the simulated and measured impedance curves (Fig. 1b) of the mechanical antenna with poled 2.5Sm-PMN-29PT material, the traditional mechanical antenna based on inverse piezoelectric effect will only work at 220 kHz. Operating at this frequency, the mechanical antenna obtains the maximum mechanical motion, thus radiating stronger EM waves compared with other frequencies. This characteristic results in a narrow resonant peak in the wireless transmission spectrum, which makes the traditional mechanical antenna seem like a narrowband antenna and may decrease the stability of communication systems. Another physical mechanism similar to the inverse piezoelectric effect, *i.e.*, electrostrictive effect, needs a lower requirement of crystallographic symmetry and the working condition. For some ferroelectric materials with a large dielectric constant, the electrostrictive strain can even reach the same or higher order of magnitude as the piezoelectric strain[15,16]. Therefore, the electrostrictive mechanical antenna is proposed to enhance the bandwidth, radiation intensity, and stability.

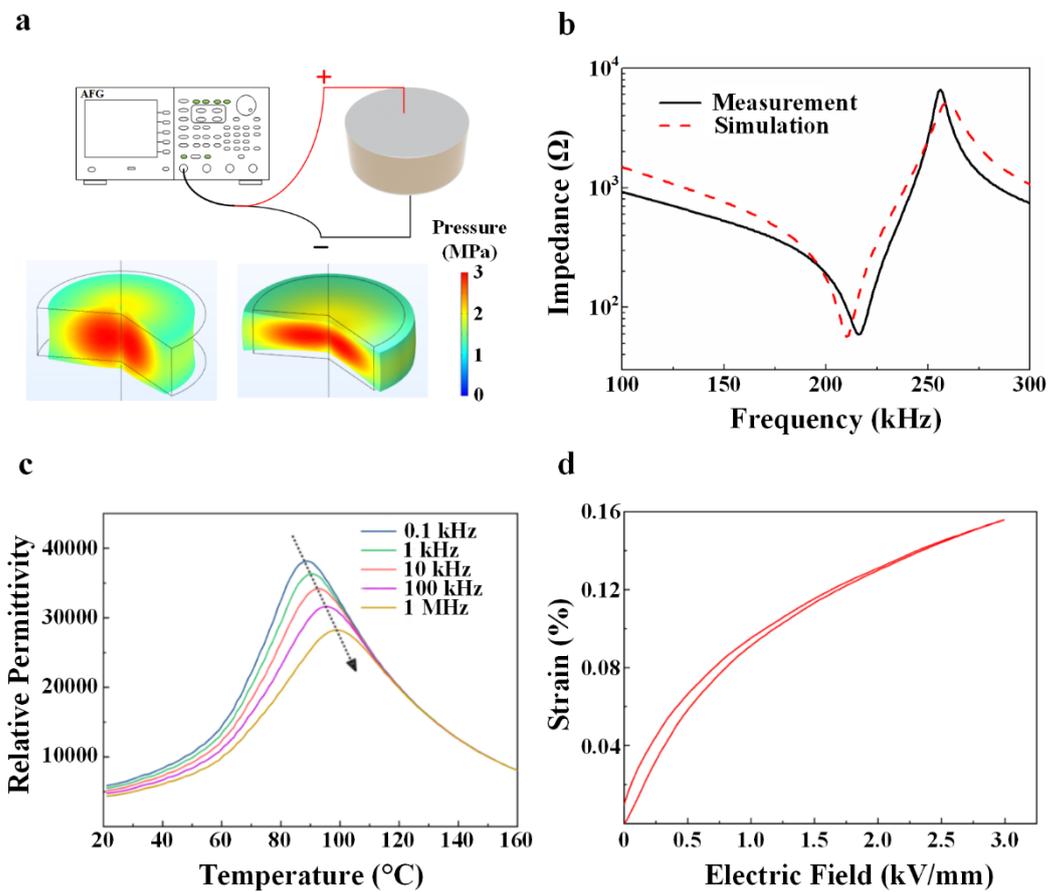

**Fig. 1| Characterization of the ceramic material. a**, Schematic diagram of the piezoelectric mechanical antenna and its **b**, impedance curves. **c**, Temperature dependence of dielectric permittivity $\varepsilon_r$ of 2.5Sm-PMN-29PT ceramic. **d**, Unipolar strain of 2.5Sm-PMN-29PT ceramic measured at different frequencies under 3 kV/mm.

A 2.5Sm-PMN-29PT ceramic chip with a thickness of 3 mm and a diameter of 8 mm is selected as the cylindrical resonator of the proposed mechanical antenna. 2.5Sm-PMN-29PT ceramics were prepared by solid-state sintering, the detailed process can be found in the Methods and elsewhere[17,18]. From the X-ray diffraction (XRD) and the X-ray energy-dispersion spectroscopy (EDS) results (Fig. S1 and Fig. S2), the ceramic sample possesses a typical perovskite phase structure and a homogeneous chemical distribution. As shown in Fig. 1c, 2.5Sm-PMN-29PT ceramic has a diffused phase transition around 90°C, which is one of the characteristics of relaxor ferroelectricity[19]. Meanwhile, it has a large $\varepsilon_r$ between room temperature and $T_c$, about 5,000 - 38,000 that is nearly an order of magnitude higher than PZT-based ceramics[20]. From the unipolar strain curves of 2.5Sm-PMN-29PT ceramic shown in Fig. 1d, the low hysteresis further verifies its ferroelectricity[21].

Mechanism of electrostriction based on the inverse piezoelectric effect and the electrostriction effect can be briefly described in Fig. 2a. The electrostrictive effect and the inverse piezoelectric effect are often confused because they produce similar physical phenomena. Actually, these two effects are distinct in nature and independent of each other. The origin of the inverse piezoelectric effect is the reorientation of the intrinsic dipole moment under an external electric field, while for the electrostrictive effect, it's the elastic displacement of ions under an applied electric field[22,23]. Different from piezoelectric strain, electrostrictive strain has no hysteresis with electric field and is not susceptible to temperature.[24] Under an external periodic electric field, ceramics expand longitudinally due to the electrostrictive effect, which leads to the periodic change of electric field intensity in the space and radiates electromagnetic waves consequently. A classical disordered slush-like domain morphology was observed from the PFM result (Fig. 2b), reflecting its significant relaxor ferroelectric property[25,26]. The SEM result (Fig. 2c) shows 2.5Sm-PMN-29PT ceramic possesses a dense microstructure with an average grain size between 5 - 8 μm.

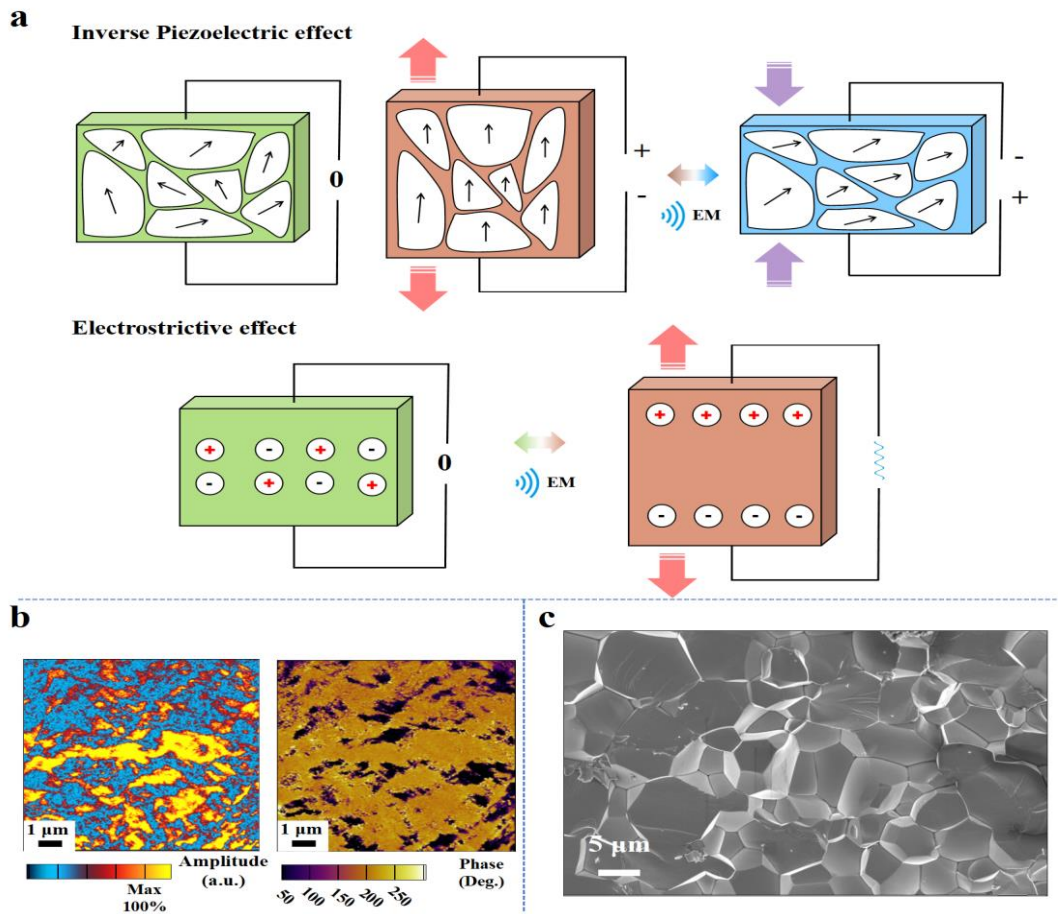

**Fig. 2| Mechanism analysis. a**, Simplified illustration of the inverse piezoelectric effect and electrostrictive effect. **b**, In-plane piezoresponse force microscopy (PFM) amplitude and phase images of 2.5Sm-PMN-29PT ceramic. **c**, Scanning electron microscopy (SEM) images of the fracture surface of 2.5Sm-PMN-29PT ceramic.

The major mechanism of the mechanical antenna in this study is detailedly discussed in the Supplementary Information. We believe that the electrostrictive strain of 2.5Sm-PMN-29PT ceramic is the main physical mechanism of its use as a cylindrical resonator in this study. Because of its different physical mechanism, the proposed mechanical antenna based on the electrostrictive effect shows excellent transmission characteristics from the traditional mechanical antennas, which will be discussed in detail below.

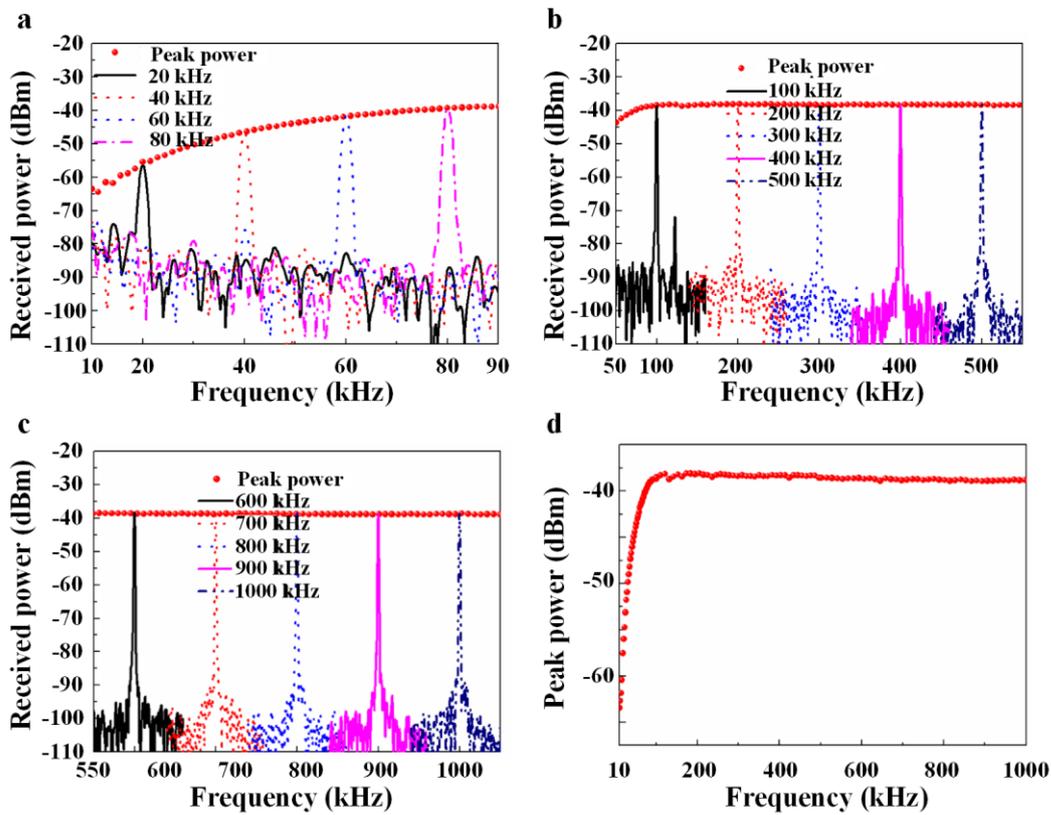

**Fig. 3| Bandwidth measurements.** Received wireless spectrum of the proposed antenna in the ranges of **a**, 10 - 90 kHz, **b**, 50 - 550 kHz, and **c**, 550 - 1050 kHz. **d**, Peak power curve.

For the frequencies below 10 kHz, the pink noise is stronger than the signal intensity and is greatly influenced by temperature. Thus, the performance of the mechanical antenna in the frequency range less than 10 kHz is not be taken into account. Using the max hold function and segmental measurement method, the continuous peak power curve in the range of 10 kHz - 1 MHz was measured. For the clear exhibition, the measured wireless spectrums with different frequencies equally distributed in all the bands were shown as examples. The displacement rate of the ions in the material increases with the frequency of feed signals under the unsaturation condition, thus enhancing the radiation intensity of the proposed antenna (Fig. 3a). In the range of 50 - 550 kHz, the received peak power reaches and stays beyond -40 dBm (Fig. 3b), which is different from the traditional piezoelectric mechanical antennas that illustrate a narrow resonant peak near the acoustic resonant frequency. In the higher frequency band of 550 kHz - 1 MHz, the received power maintains a high level and shows a slight decrement trend (Fig. 3c). In these bands, the measured wireless spectrum (Fig. 3d) of the proposed mechanical antenna obtains beyond -70 dBm peak power which can be obviously distinguished with the environmental noises. These bands thus can be used as operating frequency bands in the communication systems. Actually, the measurement range is limited by the bandwidth (20 Hz - 1 MHz) of receiving antenna used in our experiment. As expected from the trend of the measured wireless spectrum curve shown in Fig. 3d, the operating frequencies in fact include higher frequency band.

## Discussion

To investigate the influence of polarization state on the radiation performance, the proposed mechanical antenna with poled material was fabricated, and the relative measurements were also performed. It should be mentioned that all 2.5Sm-PMN-29PT ceramics used in the proposed mechanical antenna have the same parameters in composition and preparation process except for the polarization. In this condition, the measured results (Fig. S3) of the received wireless spectrum and peak power curve show coincidence with the unpoled prototype. As well as the same trend and amplitude, this mechanical antenna also exhibits ultra-wideband characteristics. These performances demonstrate that the EM radiations of these mechanical antennas are basically unaffected by the polarization state of the materials. Therefore, the mechanical antenna with unpoled material will be used in the following experiments.

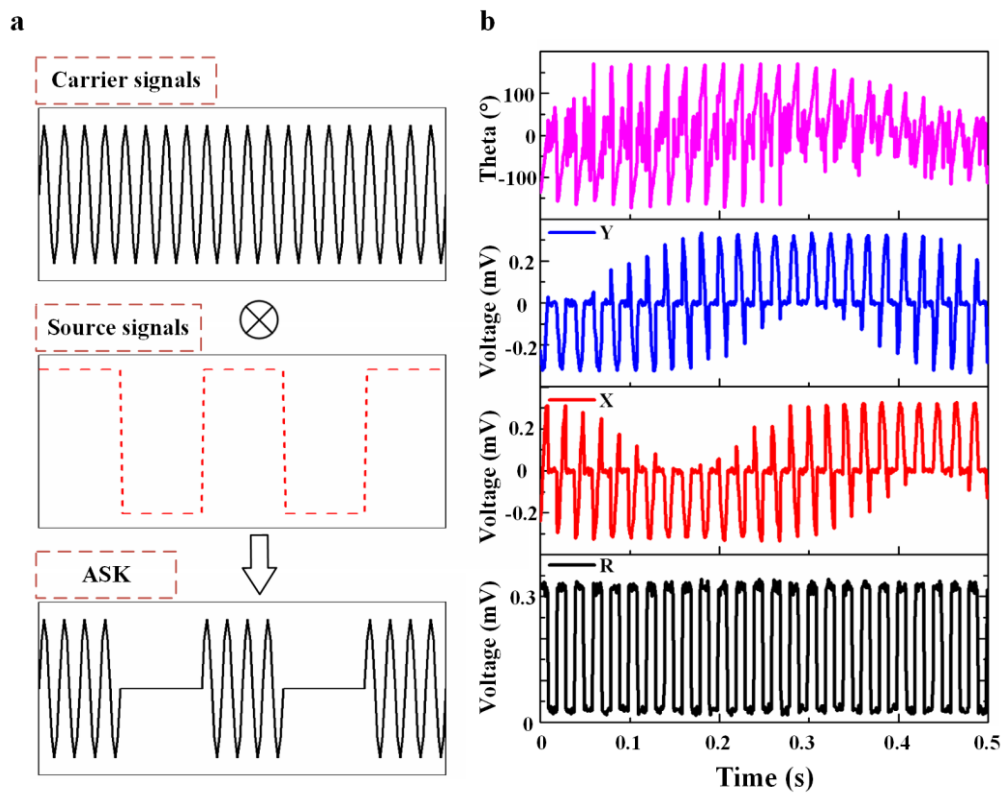

**Fig. 4| Signal modulation. a**, Schematic of ASK modulation. **b**, Received signals.

Signal modulation is a crucial part of practical communication systems. In our experimental system, a lock-in amplifier (LIA) is employed to measure the received signals under a normal noise environment. Here, amplitude shift keying (ASK) modulation scheme is used in signal processes. The frequency of input signals is modulated and amplified by an arbitrary function generator (AFG) and a power amplifier, respectively. By setting the repeat times and waveform sequence, 50% duty cycle ASK signals are generated (Fig. 4a). Here, the frequency of the carrier signals is limited by the operating frequency band of the proposed mechanical antenna, and information transmission rate is determined by the source signals' cycle that is associated with the sampling rate of the AFG and the responding rate of the LIA. The generated feed signals are directly input the positive and negative electrodes of the electrostrictive mechanical antenna without any external circuit. In the receiving terminal, the LIA with the same reference frequency as the carrier signals measures the amplitude, components, and phase of the received signals (Fig. 4b). Because of the 50% duty ratio of the source signals, the measured signals show a square waveform that represents a series of binary information. The relatively constant amplitude illustrates the stability of information transmission and low error rate. All these measured results demonstrate the feasibility of the proposed electrostrictive mechanical antenna in long-wave communication systems.

Because of the weak radiation field intensity, unable to remote transmission is an unsolved problem in current mechanical antenna design. To verify the ultra-wideband performance and remote transmission capacity of the proposed mechanical antenna, a wireless communication system is designed as shown in Fig. 5a. In low frequency communications, magnetic loop antenna is often used as the receiver to measure the magnetic flux density. Therefore, a loop antenna connected with an LIA is employed to measure the field intensity of the EM waves transmitted from the mechanical antenna in our experiment system. With this communication system, the capability of the proposed mechanical antennas in transmission distance is measured (Fig. 5b). According to the equivalent model of the dipole current, the magnetic flux density follows an inverse square relation with the distance between the transmitting antenna and the receiver. Great attenuation occurs in the vicinity near the transmitter, while the attenuation in the far-field is close to zero (the measured results of poled materials are shown in Fig. S4). The transmission distance experiments operating at 20 kHz, 200 kHz, and 1 MHz are individually performed. Compared with low frequency signals, the EM waves at the frequency of 1 MHz obtain greater transmission loss in the free space. At the frequency of 20 kHz, the mechanical antenna works under the unsaturation condition and big noise environment. Therefore, the proposed antenna operating at 200 kHz has the greatest radiation capacity. These trends are coincident with the measured spectrum (Fig. 3).

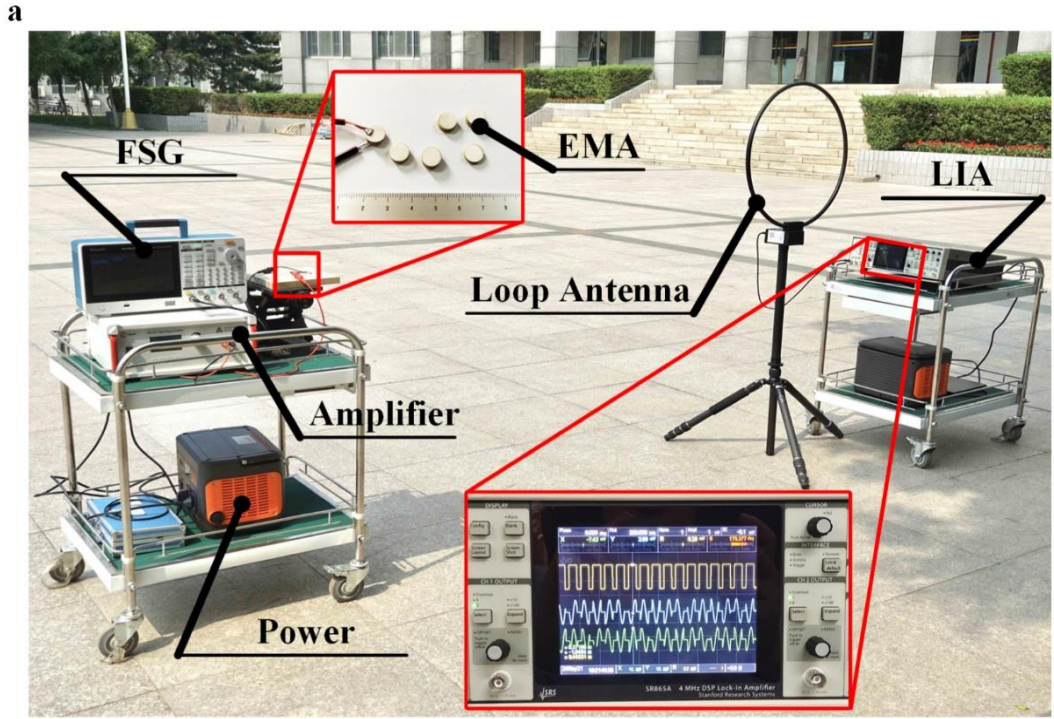

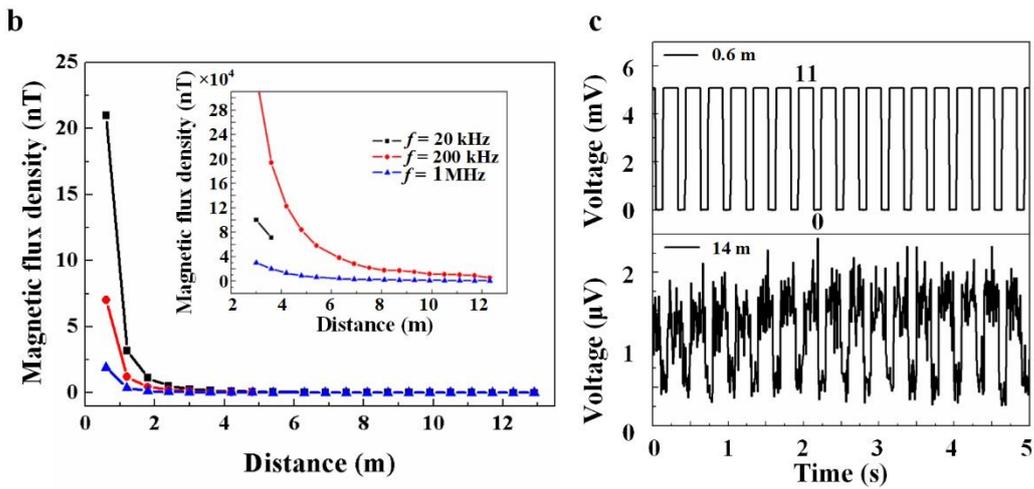

**Fig. 5| Remote transmission measurements. a**, Measurement system set-up. **b**, Magnetic flux density of the radiated EM waves versus distance. **c**, Received signals at distances of 0.6 m and 14 m.

Using the above ASK modulation at the operating frequency of 200 kHz, the different binary information of repeated '110' are transmitted to examine the ability of coding and decoding. According to the identification degree of the decoded signals, it can be found that the mechanical antenna has limit measured distance of 14 m under our experimental conditions (Fig. 5c). Moreover, the magnetic flux density at the distance of 14 m can reach the order of pT. With the help of highly precise devices and signal processes, the transmission distance will be greatly increased. Based on the relationship between the magnetic flux density and transmission distance, a fitting equation is employed to predict the magnetic flux density at greater distances. The fitting equation can be described as $y = A/(x^2)$, where $x$ represents the transmission distance, $A$ is the fitting parameter. Take the measured data at 200 kHz as an example (Fig. S5), the predicted magnetic flux density at 1 km is 8.67 fT which is efficient measure intensity for modern wireless communications. The predicted values at the other two frequencies (20 kHz and 1 MHz) are greater than 0.5 fT at the distance of 1 km. These results verify the characteristics of the proposed electrostrictive mechanical antenna in ultra-wideband and remote transmission, which provides the hardware base for long-wave communication.

In summary, we have designed an electrostrictive mechanical antenna for compact size, ultra-wide operating band, and high radiation intensity. Unlike traditional mechanical antennas which are based on inverse piezoelectric effect, the proposed mechanical antenna is mainly derived from the electrostrictive effect. Moreover, a wireless communication system was established to verify the practical performance of the proposed antenna in coding, transmitting, receiving, and decoding. Our experiment system successfully realized VLF wireless communication at the distance of 14 m.

Besides, the fitting equation of magnetic flux density predicts the remote transmission ability (beyond 1 km) of the proposed mechanical antenna. Our design will contribute to the development of long-wave antennas that may greatly reduce the volume of VLF communication systems.

## Methods

**Materials preparation.** $Pb_{0.9625}Sm_{0.025}[(Mg_{1/3}Nb_{2/3})_{0.71}Ti_{0.29}]O_3$ (2.5Sm-PMN-29PT) ceramics were fabricated by using the conventional solid-state reaction method. Firstly, the $MgNb_2O_6$ powders were prepared at 1,000 °C for 6 h by using raw materials MgO (99.9%) and $Nb_2O_5$ (99.9%). Then, the $Pb_3O_4$ (99%), $MgNb_2O_6$, $TiO_2$ (98%), and $Sm_2O_3$ (99.5%) powders were mixed by the planetary ball milling in ethanol for 24 h. The mixture was dried and calcined at 800 °C for 2 h and the synthesized powder was subjected to another round of ball milling. Before being pressed into pellets of 10 mm in diameter and 3.5 mm in thickness under 60 MPa, the obtained slurry was dried and mixed with an appropriate amount of PVA binder. Later, the PVA binder was burnout at 550 °C for 6 h. Finally, the green pellets were sintered in an alumina crucible at a temperature of 1200 °C for 3 h, buried in sacrificial powder.

**Material characterization.** The phase structure of 2.5Sm-PMN-29PT ceramics was determined by a using high-resolution X-ray diffractometer (XRD, D/max-2500V; Rigaku, Japan) with Cu Kα radiation. The surface morphology of ceramic samples was examined by using field-emission scanning electron microscopy (FE-SEM, Merlin VP compact Zeiss, Germany). A commercial atomic force microscope (MFP-3D, Asylum Research, USA) with the piezoresponse force microscopy (PFM) module was used to investigate the local ferroelectric/piezoelectric behavior of ceramic samples. Before electrical testing, both surfaces of the polished ceramic samples were painted with silver paste and then fired at 600 °C for 30 min to form electrodes. Strain $S(E)$ hysteresis loop was measured by using a ferroelectric tester (aixACCT TF Analyzer 1000, Germany). Temperature-dependent permittivity and dielectric loss were measured by using an impedance analyzer (TH2827, Changzhou Tonghui Electronic Co, China), which is coupled with a temperature-regulated sample chamber.

**Simulations.** Electromagnetic simulations were carried out with COMSOL Multiphysics, and the relative results are shown in Fig. 1a,b. Physics interfaces of piezoelectricity (solid) and electrical circuit were used to simulate the electromagnetic properties in the frequency domain. To simplify the calculations, the axisymmetric space dimension is adopted. A terminal boundary with the type of circuit, connecting to an external voltage source in the electrical circuit interface, was set to provide feed signals for the antenna. Impedance curve and stress distributions were obtained by adding relative plot groups.

**Modulation measurement.** In the experiment system, a proposed mechanical antenna and a magnetic loop antenna (SAS-565L, A.H. Systems) were used as transmitter and receiver, respectively. The mechanical antenna was driven by 20 kHz/200 kHz/1 MHz signals input directly from a power amplifier (ATA-214, Aigtek) which is connected to an AFG (AFG 31000 Series, Tektronix). The ASK modulation was performed by the control of the AFG with a peak voltage of 2.5 V. The signals from the AFG will obtain 100 times gain by using the power amplifier. In receiving terminal, an LIA (SR865A, Stanford Research Systems) with the same reference frequency as feed signals was implemented to extract the signals from extremely noisy environment. Limited by the measurement precision of the LIA, the periodic of sources signals should not be too small. For spectrum measurements, the LIA was replaced with a spectrum analyzer (N9320B, Keysight) in the experiment system.

## References

1   Marshall, R. A., Wallace, T. & Turbe, M. Finite-difference modeling of very-low-frequency propagation in the earth-ionosphere waveguide. *IEEE Trans. Antennas Propag.* **65**, 7185-7197 (2017).
2   Hua, M. et al. Very-Low-Frequency transmitters bifurcate energetic electron belt in near-earth space. *Nat. Commun.* **11**, 4847 (2020).
3   Adams, J. J. & Bernhard, J. T. A modal approach to tuning and bandwidth enhancement of an electrically small antenna. *IEEE Trans. Antennas Propag.* **59**, 1085-1092 (2011).
4   Chu, L. J. Physical limitations of omni-directional antennas. *J. Appl. Phys.* **19**, 1163-1175 (1948).
5   Stuart, H. R. & Pidwerbetsky, A. Electrically small antenna elements using negative permittivity resonators. *IEEE Trans. Antennas Propag.* **54**, 1644-1653 (2006).
6   Ziolkowski, R. W. & Erentok, A. Metamaterial-based efficient electrically small antennas. *IEEE Trans. Antennas Propag.* **54**, 2113-2130 (2006).
7   Chen, H. H. et al. Ultra-compact mechanical antennas. *Appl. Phys. Lett.* **117**, 170501 (2020).
8   Xu, J. C. et al. Metamaterial mechanical antenna for very low frequency wireless communication. *Adv. Compos. Hybrid Mater.* **4**, https://doi.org/10.1007/s42114-

021-00278-1 (2021).

9   Bickford, J. A. et al. Performance of electrically small conventional and mechanical antennas. *IEEE Trans. Antennas Propag.* **67**, 2209-2223 (2019).

10  Dong, C. Z. et al. A portable very low frequency (VLF) communication system based on acoustically actuated magnetoelectric antennas. *IEEE Antennas Wirel. Propag. Lett.* **19**, 398-402 (2020).

11  Yao, Z., Wang, Y. E., Keller, S. & Carman, G. P. Bulk acoustic wave-mediated multiferroic antennas: Architecture and performance bound. *IEEE Trans. Antennas Propag.* **63**, 3335-3344 (2015).

12  Nan, T. X. et al. Acoustically actuated ultra-compact NEMS magnetoelectric antennas. *Nat. Commun.* **8**, 296 (2017).

13  Hassanien, A. E., Breen, M., Li, M. H. & Gong, S. Acoustically driven electromagnetic radiating elements. *Sci. Rep.* **10**, 17006 (2020).

14  Kemp, M. A. et al. A high Q piezoelectric resonator as a portable VLF transmitter. *Nat. Commun.* **10**, 1715 (2019).

15  Cross, L. E., Jang, S. J., Newnham, R. E., Nomura, S. & Uchino, K. Large electrostrictive effects in relaxor ferroelectrics. *Ferroelectrics* **23**, 187-191 (1980).

16  Park, S. E. & Shrout, T. R. Ultrahigh strain and piezoelectric behavior in relaxor based ferroelectric single crystals. *J. Appl. Phys.* **82**, 1804-1811 (1997).

17  Swartz, S. L. & Shrout, T. R. Fabrication of perovskite lead magnesium niobate. *Mater. Res. Bull.* **17**, 1245-1250 (1982).

18  Li, F. et al. Ultrahigh piezoelectricity in ferroelectric ceramics by design. *Nat. Mater.* **17**, 349-354 (2018).

19  Schrettle, F., Krohns, S., Lunkenheimer, P., Brabers, V. A. M. & Loidl, A. Relaxor ferroelectricity and the freezing of short-range polar order in magnetite. *Phys. Rev. B* **83**, 195109 (2011).

20  Wu, L., Wei, C. C., Wu, T. S. & Teng, C. C. Dielectric properties of modified PZT ceramics. *J. Phys. C: Solid State Phys.* **16**, 2803-2812 (1983).

21  Lu, X. et al. Structure evolution and exceptionally ultra-low hysteresis unipolar electric field-induced strain in $(1-x)$NaNbO$_3$-$x$BaTiO$_3$ lead-free ferroelectrics. *Ceram. Int.* **44**, 5492-5499 (2018).

22  Sundar, V. & Newnham, R. E. Electrostriction and polarization. *Ferroelectrics* **135**, 431-446 (1992).

23  Newnham, R. E., Sundar, V., Yimnirun, R., Su, J. & Zhang, Q. M. Electrostriction: Nonlinear electromechanical coupling in solid dielectrics. *J. Phys. Chem. B* **101**, 10141-10150 (1997).

24  Chen, B. et al. Large electrostrictive response in lead halide perovskites. *Nat. Mater.* **17**, 1020-1026 (2018).

25  Takenaka, H., Grinberg, I., Liu, S. & Rappe, A. M. Slush-like polar structures in single-crystal relaxors. *Nature* **546**, 391-395 (2017).

26  Kumar, A. et al. Atomic-resolution electron microscopy of nanoscale local structure in lead-based relaxor ferroelectrics. *Nat. Mater.* **20**, 62–67 (2021).

## Supporting Information

**Supplementary Note 1. Selection of materials**

Typical ferroelectric materials, such as lead zirconate titanate (PZT) and lithium Niobate (LN), exhibit a sharp dielectric peak at the ferroelectric-paraelectric phase transition temperature (*i.e.*, Curie temperature $T_c$). Below this temperature, the material has an inverse piezoelectric effect, while the inverse piezoelectric effect disappears above this temperature. It's known that much heat will be accumulated during the piezoelectric materials work at a periodic electric field, especially at high frequency. Therefore, the mechanical antennas based on the inverse piezoelectric effect of ferroelectric materials may not suitable for long-term operation under high-frequency fields. Moreover, piezoelectric materials such as PZT and LN are not ideal materials for mechanical antennas due to the hysteresis effect of field-induced strain. For typical ferroelectric materials, ferroelectric domain switching makes a decisive contribution to the total strain, which leads to the hysteresis effect of the field-induced strain meanwhile. Generally, the more flexible the ferroelectric domain switch, the more serious the hysteresis effect will be.

Different from the typical ferroelectric materials, relaxor ferroelectric materials have a diffused phase transition, thus its dielectric constant can maintain a high value over a wide temperature range near the phase transition temperature. When the temperature is constant, the strain $S_E$ caused by the electrostrictive effect and the polarization intensity $P$ can be represented as

$$S_E = QP^2 = Q(\varepsilon_r\varepsilon_0 E)^2 \qquad (S1)$$

where $Q$ is the electrostriction coefficient, $E$ is the applied electric intensity, $\varepsilon_r$ and $\varepsilon_0$ are relative dielectric constant and dielectric constant of vacuum respectively. Therefore, the relaxor ferroelectric materials tend to obtain a large electrostrictive strain near the phase transition temperature since they have a huge and stable $\varepsilon_r$ at this temperature range. Furthermore, the electrostrictive strain has no hysteresis effect and do not affected by the $T_c$, because this strain does not come from the ferroelectric domain switching and the asymmetry of lattice. PMN-PT is a classical relaxor ferroelectric material, and its $T_c$ is lower than the room temperature. By doping with 2.5 mol% Sm, Li et al enhanced the $T_c$ of PMN-29PT ceramic to 80 – 90 °C and obtained a huge $d_{33}$ (1,400 – 1,500 pC/N) and $\varepsilon_r$ (13,000) around room temperature[1]. The typical relaxor ferroelectric property and the excellent dielectric property make 2.5Sm-PMN-29PT ceramics a potential candidate for mechanical antennas based on electrostrictive effect.

**Supplementary Note 2. Discussion on mechanism**

As mentioned above, the electrostrictive strain $S_E$ is proportional to the square of $\varepsilon_r$. Therefore, the electrostrictive strain in relaxor ferroelectric materials can reach the same order of magnitude as the piezoelectric strain, which has been widely studied in previous works[2,3]. Generally, it's believed that the piezoelectric effect and the electrostrictive effect contribute simultaneously to the total field-induced strain in relaxor ferroelectric material below the $T_c$, while the piezoelectric strain will gradually vanish when temperature is close to the $T_c$. Moreover, the piezoelectric strain consists of two parts, *i.e.*, lattice distortion and domain wall switching. The threshold electric field required for domain wall switching is in the kV/mm scale, thus the piezoelectric strain at low electric field mainly comes from the lattice distortion. For this part of the piezoelectric strain, the intrinsic contribution is linearly related to the electric field.

In this work, 2.5Sm-PMN-29PT ceramic exhibited significant relaxor ferroelectric property, as demonstrated in Fig. 1c,d, and 2b. During the experiment, the temperature of the material was enhanced to close to the $T_c$ under an applied high-frequency electric field above MHz, but its transmission performance kept constant. In addition, the maximum amplitude of the applied electric field in this experiment is 250 V, which is much lower than the threshold electric field of the ferroelectric domain switching. Based on these findings, we believe that the electrostrictive effect is the main mechanism of the mechanical antenna proposed in this study.

**Supplementary Figures**

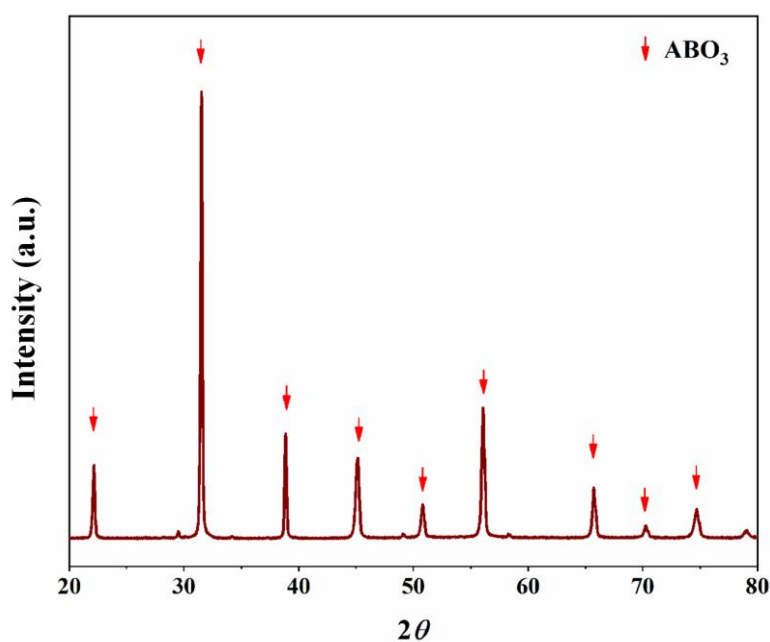

**Supplementary Figure S1| X-ray diffraction (XRD) patterns of 2.5Sm-PMN-29PT ceramic**. A high-resolution X-ray diffractometer (XRD, D/max-2500V; Rigaku, Japan) with Cu Kα radiation was utilized to determine its phase structure. A typical perovskite phase was observed and the characteristic diffraction peaks indicate a rhombohedral phase structure, which corresponds with the previous work[1].

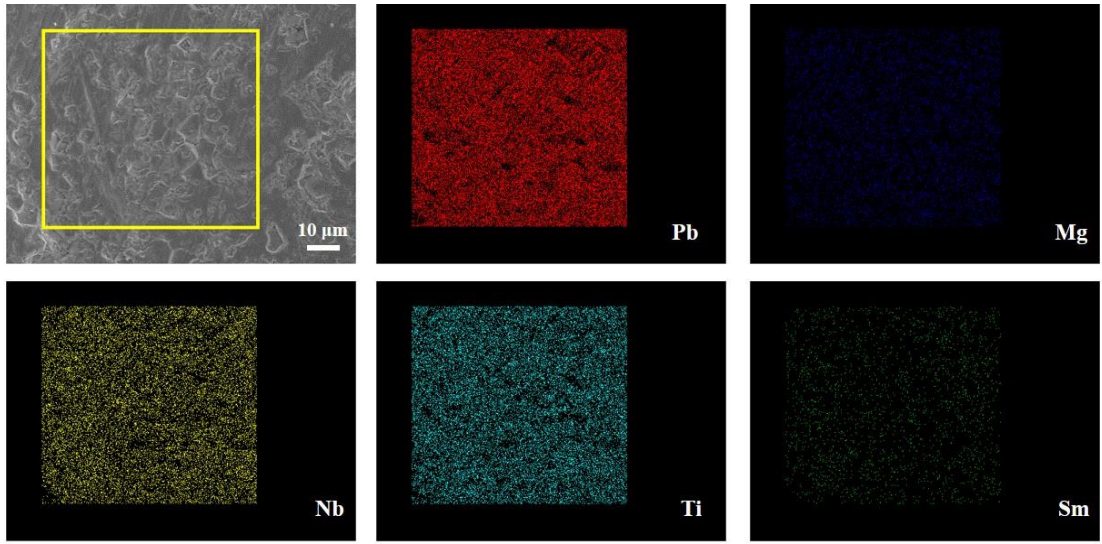

**Supplementary Figure S2| Energy dispersive spectroscopy (EDS) results of the 2.5Sm-PMN-29PT ceramic cross-section.** Distributions of elements were examined by using field-emission scanning electron microscopy (FE-SEM, Merlin VP compact Zeiss, Germany). All elements are homogeneously distributed, and no element enrichment is observed.

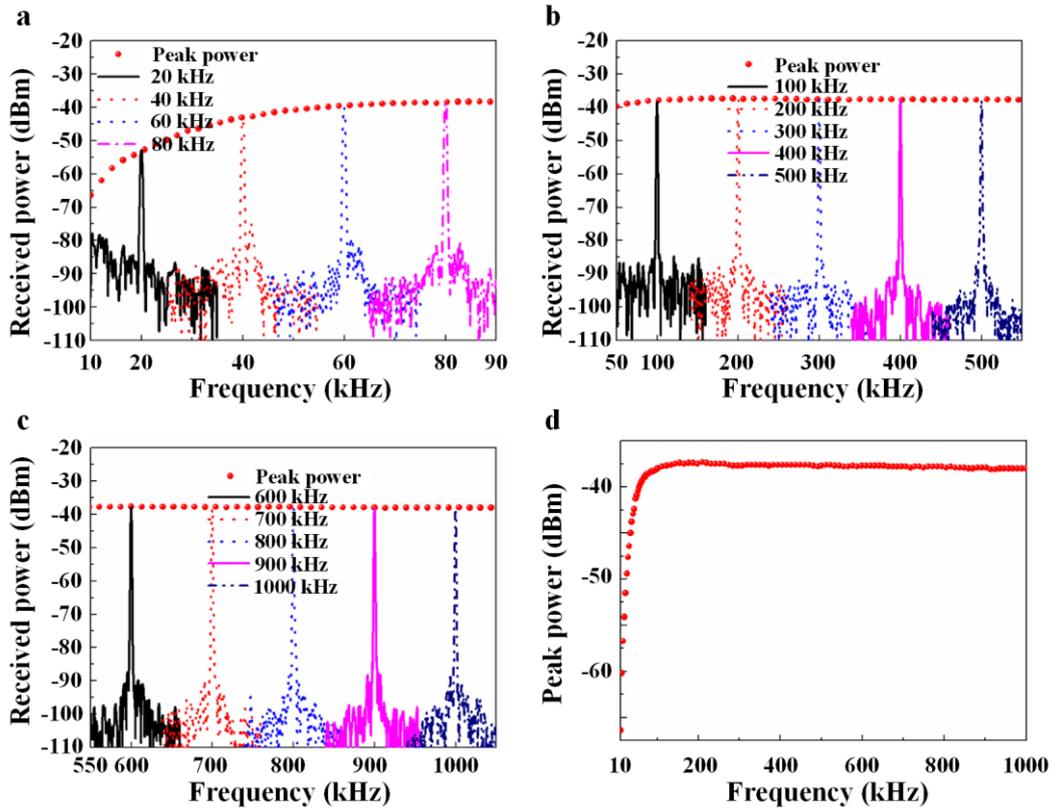

**Supplementary Figure S3| Bandwidth of the proposed mechanical antenna with poled materials. a**, Received wireless spectrum of the proposed antenna in the ranges of 10 - 90 kHz, **b**, 50 - 550 kHz, and **c**, 550 - 1050 kHz. **d**, Peak power curve.

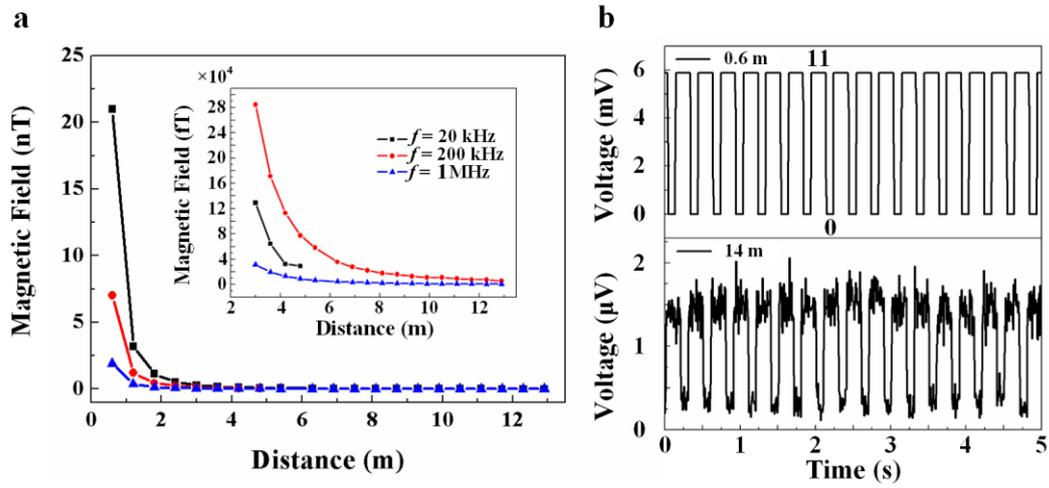

**Supplementary Figure S4| Remote transmission measurements of the mechanical antenna with poled materials. a**, Magnetic flux density of the radiated EM waves versus distance. **b**, Received signals at distances of 0.6 m and 14 m.

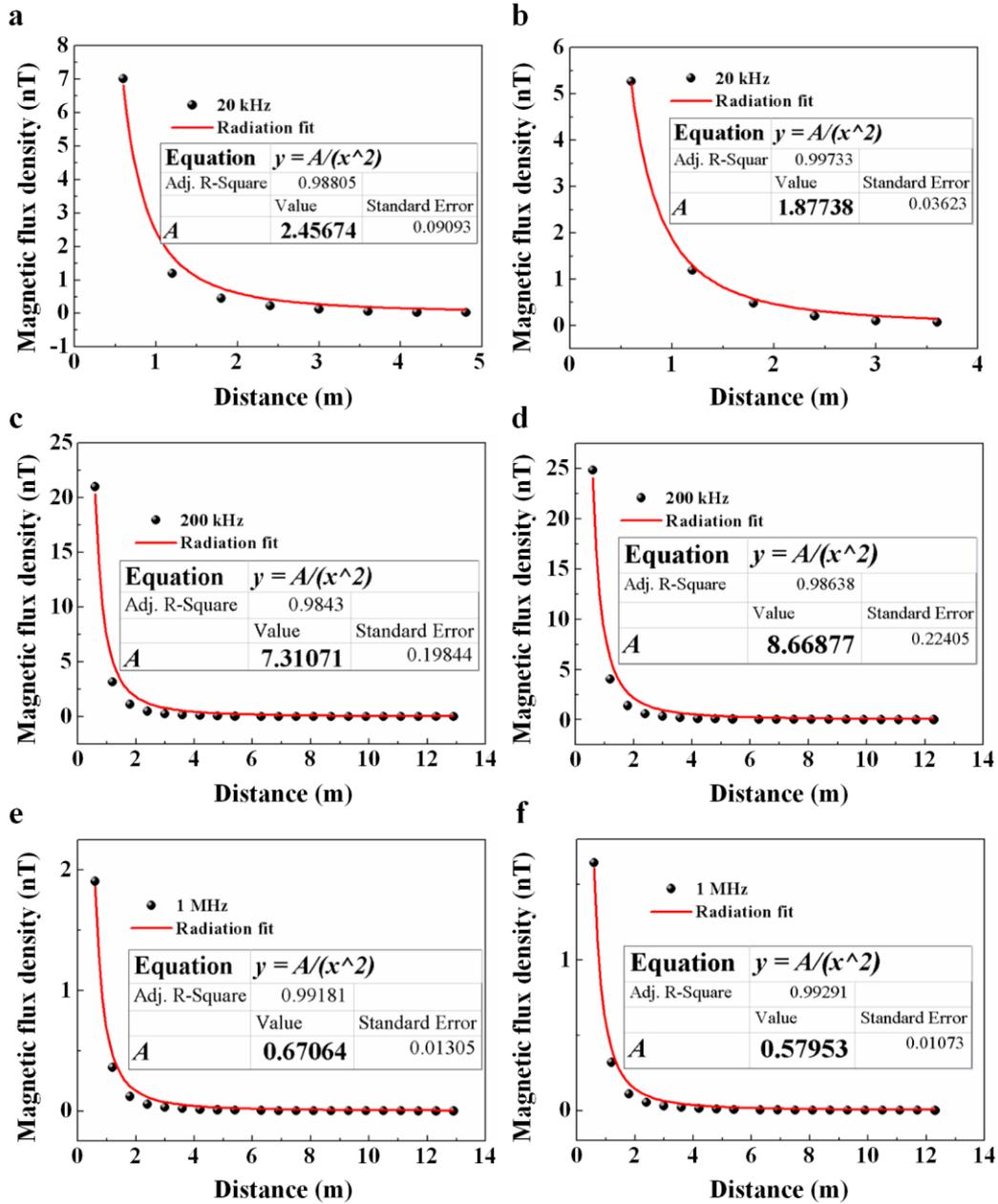

**Supplementary Figure S5| Predictions of the magnetic flux density at further distances.** Magnetic flux density fitting curves of the proposed mechanical antennas with poled and unpoled materials at **a, b** 20 kHz, **c, d** 200 kHz, and **e, f** 1 MHz. According to the fitting equations, the magnetic flux density in the distance can be predicted. For instance, the predicted

values at 1 km away from the proposed antenna with unpoled material are respectively 1.88 fT, 8.67 fT, and 0.58 fT at 20 kHz, 200 kHz, and 1 MHz. These values are in a testable range under current receiving technology, which demonstrates the remote transmission ability of the proposed ultra-band electrostrictive mechanical antenna within the bandwidth.

# References

1    Li, F. et al. Ultrahigh piezoelectricity in ferroelectric ceramics by design. *Nat. Mater.* **17**, 349-354 (2018).
2    Cross, L. E., Jang, S. J., Newnham, R. E., Nomura, S. & Uchino, K. Large electrostrictive effects in relaxor ferroelectrics. *Ferroelectrics* **23**, 187-191 (1980).
3    Park, S. E. & Shrout, T. R. Ultrahigh strain and piezoelectric behavior in relaxor based ferroelectric single crystals. *J. Appl. Phys.* **82**, 1804-1811 (1997).